\documentclass[10pt,prl,aps,twocolumn,floatfix,showpacs]{revtex4}
\usepackage{graphicx}
\usepackage{amssymb}
\usepackage{bm}% bold math

\begin{document}

\title{Quantum Monte Carlo simulations of bosonic and fermionic impurities  \\
in a two-dimensional hard-core boson system}

\author{Anders W. Sandvik} 
\affiliation{Department of Physics, Boston University, 
590 Commonwealth Avenue, Boston, Massachusetts 02215}

\date{\today}

\begin{abstract}
A two-dimensional lattice hard-core boson system with a small fraction of bosonic or fermionic 
impurity particles is studied. The impurities have the same hopping and interactions as the 
dominant bosons and their effects are solely due to quantum statistics. Quantum Monte Carlo 
simulations are carried out in which paths of the dominant boson species are sampled and a summation 
is performed over all second-species paths compatible with the permutation cycles. Both kinds of 
impurities reduce modestly and equally the Kosterliz-Thouless superfluid transition temperature. 
However, the effective impurity interactions are found to be qualitatively different at lower 
temperatures; fermions are repulsive and further suppress superfluidity as $T \to 0$.
\end{abstract}


\maketitle

Ultracold atoms in optical lattices offer unprecedented opportunities to realize novel quantum 
states of matter \cite{bloch}. One prospect is to tailor systems to mimic hamiltonians 
of fundamental interest in condensed matter physics, e.g., the Hubbard model \cite{hofstetter}. 
Another interesting route is to explore states that do not have any realizations in naturally occurring 
systems. The use of mixtures of different atomic species open up almost endless possibilities. In the 
case of two species, there are three mixture classes; bose-bose, fermi-fermi, and bose-fermi, 
with the latter perhaps offering the most interesting prospects. Several exotic phases have been predicted 
theoretically, e.g., supersolids \cite{buchler}, several different Mott states \cite{sengupta}, multiply 
degenerate quantum-disordered states with glass-like properties \cite{albus}, paired states with various 
orbital symmetry \cite{mathey}, and a host of states of effective fermion-boson composite particles 
\cite{lewenstein}. Experimentally, there are intriguing results indicating a strong influence of a small 
admixture of fermions in boson gases in optical lattices \cite{exp}.

In this Letter a simple two-species model will be considered in which all particles have identical
hoppings and interactions, posing a clean way to elucidate the fundamental role of quantum statistics. 
The situation can be realized experimentally in systems with two isotopes of the same atom, e.g., 
$^{\rm 6}$Li-$^{\rm 7}$Li mixtures \cite{truscott}. A 1D model of this kind has been studied using various 
analytical approaches \cite{imambekov}. Here a quantum Monte Carlo (QMC) method is developed for a low 
concentration of fermionic or bosonic impurities in a bath of bosons. The scheme is applied to a 2D model. 
Defining creation operators $a^\dagger_i$ and $b^\dagger_i$ for the two species the hamiltonian is 
\begin{equation}
H=-t\sum_{\langle ij\rangle} (a^\dagger_i a_j + a^\dagger_j a_i +
b^\dagger_i b_j + b^\dagger_j b_i),
\label{ham}
\end{equation}
where $\langle ij\rangle$ denotes nearest neighbors on a square lattice of $N=L \times L$ sites  
with periodic boundaries. Species A is bosonic, whereas the $B$ particles can be either bosons or fermions. 
Both are subject to a hard-core constraint, i.e., the sites are either empty or singly-occupied. 
Half-filling in the canonical ensemble will be considered here; $n = n_A + n_B = N/2$, 
where $n_A=\sum a^\dagger_i a_i$ and $n_B=\sum b^\dagger_i b_i$. The trapping potential necessary 
to model experiments is left out, but can easily be incorporated in future work using the QMC 
scheme introduced below. 

With a fermionic $B$ species, the hamiltonian as a function of $n_B/n$ 
interpolates between spinless fermions and the standard hard-core boson model, which is equivalent to 
the $S=1/2$ XY model and undergoes a Kosterliz-Thouless (KT) transition to a superfluid with power-law 
off-diagonal correlations at temperature 
$T_{\rm KT}/t \approx  0.69$ \cite{tkt}; at $T=0$ it is long-range ordered. The spinless fermion 
model, on the other hand, has a metallic ground state and does not undergo any finite-$T$ transition. 
It is then interesting to consider the evolution of the ground state and finite-$T$ properties 
as a function of the fermion fraction. Here the A species will be considered dominant, the $B$ 
particles acting as impurities; $n_B \ll n_A$.

The purely bosonic ground state does not change when some A particles are replaced by another 
boson species. The excitations are affected, however, and one can expect a reduction in $T_{\rm KT}$ 
relative to the single-species system. It will be shown here that the effective interaction between 
the impurities is attractive at high temperatures but changes, in a singular way, to repulsive at 
$T_{\rm KT}$. Bosonic impurities become attractive at lower $T$, whereas fermions stay repulsive as 
$T \to 0$. The effects of the impurities on the KT transition are independent of their statistics, 
but there are indications of a more dramatic suppression of superfluidity by fermions as $T \to 0$.

{\it QMC Algorithm.}---Consider finite-$T$ QMC methods in which the density matrix ${\rm e}^{-H/T}$ 
($k_B=1$) is written as a sum of operators $P_i$ which propagate real-space states  
$|\alpha \rangle$ such that $P_i |\alpha \rangle = W_i(\alpha)|\alpha\rangle$. The paths $(i,\alpha)$ 
are importance-sampled according to their weights $W_i(\alpha)$ in the partition function 
$Z=\sum_{i,\alpha}W_i(\alpha)$. Such path-integral methods based on "time-slicing" can be formulated 
in the continuum \cite{ceperley} and on lattices \cite{worldline}. In recent years very efficient algorithms 
for updating the paths have been devised \cite{prokofev,directed,boninsegni} and systems with thousands of 
bosons can now be routinely simulated even at low temperatures. On the lattice, the alternative stochastic 
series expansion (SSE) 
approach \cite{sse,sseprb,directed}, in which ${\rm e}^{-H/T}$ is Taylor expanded to all contributing orders, 
is often more efficient and will be employed here. The details of the method are unimportant, 
however, and the scheme for treating impurity particles outlined below should be applicable with any 
standard boson QMC algorithm. 

\begin{figure}
\includegraphics[width=3.8cm, clip]{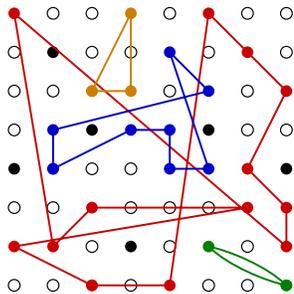}
\caption{(Color online) Permutation cycles for an $8\times 8$ system at $T/t=0.5$. 
Open and solid circles represent empty sites and particles, respectively. The net effect 
of the SSE propagation of the state is to cyclically permute the particles connected by the 
closed paths. The isolated solid (black) circles are particles not undergoing any net permutations; 
they are considered permutation cycles of length one.} 
\label{fig1}
\vskip-3mm
\end{figure}

Identical particles in $P_i |\alpha \rangle$ are permutations of those in $|\alpha\rangle$. 
The permutations can be decomposed into cycles $C_i$ in which $m_i$ particles are 
permuted independently of the other particles. An example from an actual SSE simulation 
of an $8\times 8$ lattice with $32$ bosons is shown in Fig.~\ref{fig1}. In the case of 
a multi-species system, a term (path) contributes to the partition function as long as all 
particles within each individual permutation cycle are identical; particles in different 
cycles do not have to be identical. Now, if all the interaction and hopping parameters 
are identical for all species, the weight $W_i(\alpha)$ is independent of the (allowed)
distributions of the particles over the cycles. A simulation can then be carried out for all 
identical hard-core bosons ($A$ particles) and, subsequently, when measuring observables, two 
(or more) species can be considered by substituting all the particles in some of the cycles with 
$B$ particles. All ways of filling the cycles can be summed up exactly, 
including fermionic signs.

Denote by $N_c(m)$ the number of cycles of length $m$ and by $n_c(m)$ the number of cycles 
filled with $B$ particles. Then the total number of ways of distributing $n_b$ bosons of type 
$B$ in the system (substituting the same number of $A$ particles) is given by
\begin{equation}
w_b = \sum_{\{ n_{\rm c}\}} \prod_{m=1}^{n_b} {N_{\rm c}(m) \choose n_{\rm c}(m)},
\label{wb}
\end{equation}
where, for a fixed number $n_b$ of B particles, the sum is over all cycle fillings satisfying the 
constraint $\sum_{m} mn_{\rm c}(m)=n_b$.
Thus, if the paths are importance-sampled using the weight $W_i(\alpha)$, the measurements of 
some observable $\hat O$ should be weighted by $w_b$;
\begin{equation}
\langle \hat O \rangle = \frac{\sum_{i,\alpha}w_b(i,\alpha)\langle O_{i,\alpha}\rangle}
{\sum_{i,\alpha}w_b(i,\alpha)}.
\end{equation}
Here the sums are over the paths actually sampled in the simulation.
The estimator $\langle O_{i,\alpha}\rangle$ is an average over the $w_b(i,\alpha)$ different
cycle fillings. 

\begin{figure}
\includegraphics[width=6.5cm, clip]{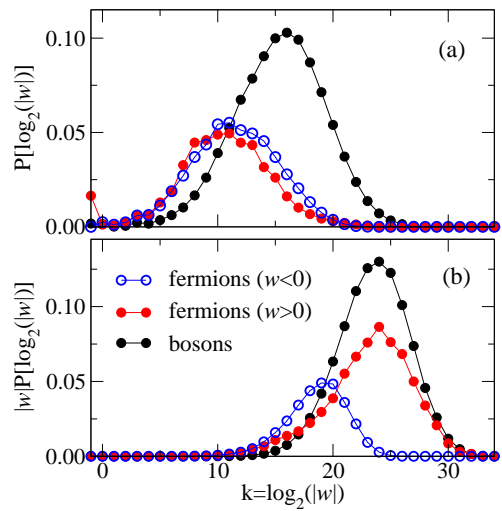}
\caption{(Color online) (a) Weight distribution for fermions ($w_f \ge 0$ and $w_f < 0$ separately) and 
bosons in simulations of an $L=16$ lattice with $n_{b,f}=12$ at $T/t=0.25$. Bin $k$ (integer) represents the 
probability of $k \le \log_{\rm 2}(|w|) < k+1$ (the special case $w=0$ is in the $k=-1$ bin). (b) The 
probability times the weight, i.e., the actual contribution of terms in a given weight range (normalized to 
unity for both bosons and fermions). The scheme breaks down when a significant fraction of the weight in a 
histogram in (b) extends far into the rarely sampled right tail of the probability histogram in (a). Here 
the average fermionic sign, i.e., the ratio of fermionic and bosonic weights, $\langle S \rangle = 
\langle w_f\rangle/\langle w_b\rangle \approx 0.03$. The effective sign in the simulation is the difference 
between the $w_f \ge 0$ and $w_f < 0$ histograms in (b) and is larger; $\langle S\rangle_{\rm eff} \approx 0.4$} 
\label{fig2}
\vskip-3mm
\end{figure}

In the case of fermions, each cyclic permutation of an even number of particles yields
a minus sign. Thus, in the presence of $n_f$ fermionic B particles the weight is
\begin{equation}
w_f = \sum_{\{ n_{\rm c}\}} \prod_{m=1}^{n_f} {N_{\rm c}(m) \choose n_{\rm c}(m)}
(-1)^{(m-1)n_{\rm c}(m)},
\label{wf}
\end{equation}
with the constraint $\sum_{m} mn_{\rm c}(m)=n_f$.

It will be demonstrated that it is feasible to evaluate exactly the 
weights (\ref{wb}),(\ref{wf}) even for a relative large number of impurity particles---results 
will be presented for $n_{b,f} \le 32$ on a $32 \times 32$ lattice. 
For some observables the estimator $O_{i,\alpha}$ is independent of the cycle filling whereas 
in other cases the evaluation of its average may be more complicated. In some cases, it may be 
necessarily to carry out a separate sampling of the average estimator for each path. Here only 
the internal energy and the phase stiffness will be considered, both of which have SSE estimators 
involving operator counts in $P_i$, which are independent of the distribution of $A$ and $B$ particles. 
The energy $E=\langle H\rangle$ is given simply by the average order $p$ of the SSE Taylor expansion; 
$E = - \langle p\rangle/\beta$ \cite{sse}. The stiffness is the second derivative of the energy 
$E(\phi)$ with respect to a twist $\phi$ in the boundary condition. It is obtained by averaging 
the squared winding number \cite{pollock,sseprb}.

A limitation of the cycle summation approach is that the relative fluctuations in the weighs $w_{b,f}$
grow as $n_{b,f}$ increases. Then only a decreasing subset of the generated configurations will contribute
significantly to computed quantities. In addition, for fermions there is still a sign problem, although
the summation over many cycle fillings with different signs does alleviate it significantly. These issues are 
illustrated and further discussed in Fig.~\ref{fig2}. Another problem is that it becomes prohibitively time 
consuming to exactly evaluate $w_{b,f}$ for large $n_{b,f}$. For moderate $n_{b,f}$, it is possible to 
obtain results on large lattices in temperatures regimes where interesting physics takes place, as will be shown
next.

\begin{figure}
\includegraphics[width=6.25cm, clip]{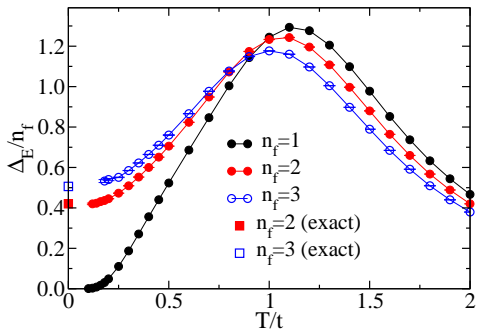}
\caption{(Color online) QMC and Lanczos results for the internal energy of a $4 \times 4$ system with 
$n_f=1,2,3$ fermions, relative to the energy of the purely bosonic system.} 
\label{fig3}
\vskip-3mm
\end{figure}

{\it Results.}---The 
correctness of the QMC scheme was confirmed by exact diagonalization results for a $2\times 4$ system at 
finite $T$. Even using momentum conservation, it is difficult to completely diagonalize 
a $4\times 4$ system with two particle species. The ground state can be obtained using the Lanczos method, 
however. 

An interesting quantity is the change $\Delta_E(n_{b,f})$ in the internal energy relative to the
$n_{b,f}=0$ energy; $\Delta_E = E(n_{b,f})-E(0)$. Fig.~\ref{fig3} shows how QMC results for $\Delta E(n_f)/n_f$ 
approach corresponding Lanczos results for $n_f=1,2,3$. Note that with $n_f=1$, fermion anticommutation does 
not come into play and the same bosonic ground state as with $n_f=0$ is obtained as $T \to 0$.

\begin{figure}
\includegraphics[width=6.7cm, clip]{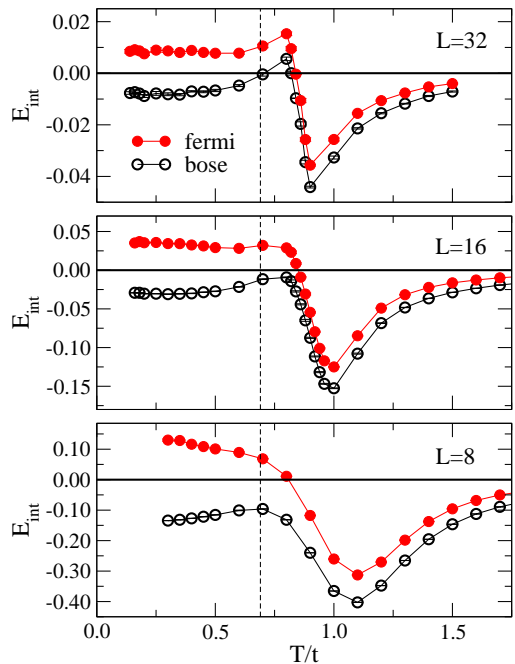}~~
\caption{(Color online) Interaction energy for $8$ impurities in $L=8,16$, and $32$ lattices.
The dashed lines indicate $T_{\rm KT}$.} 
\label{fig4}
\vskip-3mm
\end{figure}

An effective interaction energy between impurities can be defined by subtracting the single-impurity
result $\Delta_E(1)$; $E_{\rm int} = \Delta_E(n_{b,f})/n_{b,f} - \Delta_E(1)$. This quantity is shown in 
Fig.~\ref{fig4} for $8$ bosonic and fermionic impurities in lattices of size $L=8,16$, and $32$. At high 
temperatures $E_{\rm int} < 0$, indicating effectively attractive interactions for both bosons and fermions. 
A singular behavior appears to develop at $T\approx T_{\rm KT}$, where for large systems both the fermionic and 
bosonic interactions first become increasingly attractive and then suddenly turn repulsive. At lower $T$ the 
fermionic interactions stay repulsive while the bosonic ones become attractive again. The behavior is very similar 
for all impurity numbers $n_f \ge 2$ studied. In the case of fermions (bosons) $E_{\rm int}$ at low $T$ increases 
as a function of $n_f$ (decreases as a function of $n_b$), as would be expected for repulsive (attractive) 
interactions. When negative, $E_{\rm int}$ is seen to decrease as a function of the system size, indicating 
that the impurities do not demix but remain distributed throughout the system. 

Note that $n_f$ is fixed in Fig.~\ref{fig4}, i.e., the impurity concentration decreases with increasing $L$. 
If the impurities do not segregate, it is clear that anticommutation will not be significant down to some 
concentration dependent $T$ at which fermions, at their typical separation, can begin to permute. Thus, the 
behavior around $T_{\rm KT}$ for a low impurity density should be a bosonic feature. The qualitative 
difference between bosonic and fermionic impurities emerges at lower temperature, where the fermions remain 
repulsive down to $T \to 0$ whereas the bosons become attractive [and eventually the effective interactions vanish 
as $E(n_b>0) \to E(0)$ for bosons].

It appears likely that the singularity seen developing in Fig.~\ref{fig4} should move exactly to $T_{\rm KT}$ 
as $L \to \infty$ and, hence, that vortices play a decisive role in the effective interactions. One can speculate 
that the each impurity particle associates with a vortex. The low-$T$ repulsive fermionic interactions may then 
point to vortex-antivortex pairs that increase in size and do not annihilate. There would then be 
$n_f/2$ vortex-antivortex pairs left at $T=0$.

\begin{figure}
\includegraphics[width=6.75cm, clip]{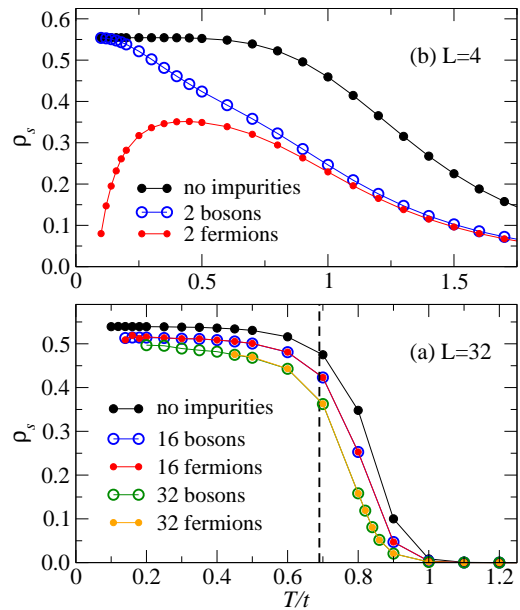}
\caption{(Color online) Phase stiffness for (a) $L=4$ systems with $n_{b,f}=0,2$ and (b) $L=32$ systems
with $n_{b,f}=0,16,32$.} 
\label{fig5}
\vskip-3mm
\end{figure}

The influence of the bosonic and fermionic impurities on the phase stiffness (the superfluid density 
\cite{pollock}) around $T_{\rm KT}$
is almost indistinguishable for large system sizes, as shown for $L=32$ in Fig.~\ref{fig5}(a). 
In both cases the stiffness is mildly suppressed, pointing to a modest reduction in $T_{\rm KT}$ for 
impurity concentrations $1/32$ ($16$ impurities) and $1/16$ ($32$ impurities). However, as shown in 
Fig.~\ref{fig5}(b), there are significant differences in the case of two impurities in a $4\times 4$
lattice at low temperatures. As expected, for bosonic impurities the stiffness approaches that of the 
single-species system. Two fermions, on the other hand, strongly suppress the stiffness as $T \to 0$. In fact, 
$\rho_s \to - \infty$ as $T \to 0$, which can be traced to the minimum in the energy $E(\phi)$ as a function 
of the boundary twist $\phi$ being slightly away from $\phi =0$. In addition to this anomaly, the ground 
state does not have momentum $q=0$ but is degenerate with ${\bf q}=(\pm \pi/2,0),(0,\pm \pi/2)$. These 
features suggest that the ground state is frustrated, which may again be indicative of the two fermions
being tied to a vortex-antivortex pair. A drop in the stiffness is also seen for a small ($2-4$) number of 
fermions in larger lattices, at a $T$ which decreases as $L$ is increased for fixed $n_f$. The ground 
state may thus be insulating.

{\it Summary and Conclusions.}---The 
QMC method developed here enables studies of bosonic as well as fermionic impurity 
particles in a bath of hard-core bosons. Here a system with no interactions apart from the hard-core
constraint was studied, but the method is applicable also in the presence of other interactions as long
as they are equal for all species (inter- and intra-species). 

In the model studied here, signs of a singular effective impurity interaction are 
seen at $T_{\rm KT}$, pointing to a decisive role of vortices. At low $T$ the 
effective fermionic interactions are repulsive whereas the bosonic ones are attractive.
A scenario suggested by these findings is that impurities associate with vortices. An effectively repulsive 
fermionic impurity interaction, along with a drop observed in the phase stiffness, would then imply that 
some vortex-antivortex pairs remain as $T \to 0$, potentially leading to an insulating ground state.
For bosonic impurities, the ground state is the same as without impurities. 
Future studies will address correlations in systems with a very small number of impurities at
lower temperatures. This should give further insights into the nature of the ground state
of the bose-fermi mixture.

In the presence of a trapping potential, the repulsive fermions should be expelled to the condensate boundary at
low temperatures. However, simulations of the grand-canonical ensemble, but with fixed $n_f$, demonstrate
an effectively attractive fermi-bose interactions (the filling $\langle n \rangle$ becomes larger than 
$1/2$). Thus the boundary should be in an interesting non-trivial mixed state. 

I would like to thank Anatoli Polkovnikov and Asle Sudb$\o$ for useful discussions.
This work was supported by the NSF under grant No.~DMR-0513930.

\null


\vskip-9mm
\begin{thebibliography}{00}

\bibitem{bloch}
I. Bloch, J. Dalibard, and W. Zwerger, arXiv:0704.3011 (to appear in Rev. Mod. Phys.).

\bibitem{hofstetter}
W. Hofstetter, J. I. Cirac, P. Zoller, E. Demler, and M. D. Lukin,
Phys. Rev. Lett. {\bf 89}, 220407 (2002).

\bibitem{buchler}
H. P. B\"uchler and G. Blatter, Phys. Rev. Lett. {\bf 91}, 130404 (2003).

\bibitem{sengupta}
K. Sengupta, N. Dupuis, and P. Majumdar, Phys. Rev. A {\bf 75}, 063625 (2007).

\bibitem{albus}
A. Albus, F. Illuminati, and J. Eisert, Phys. Rev. A {\bf 68}, 023606 (2003).

\bibitem{mathey}
L. Mathey, S.-W. Tsai, and A. H. Castro Neto, Phys. Rev. Lett. {\bf 97}, 030601 (2006).

\bibitem{lewenstein}
M. Lewenstein, L. Santos, M. A. Baranov, and H. Fehrmann, Phys. Rev. Lett. {\bf 92}, 050401 (2004).

\bibitem{exp}
K. G\"unter {\it et al.}, Phys. Rev. Lett. {\bf 96}, 180402 (2006); S. Ospelkaus {\it et al.},
{\it ibid}, {\bf 96}, 180403 (2007).

\bibitem{truscott}
A. Truscott {\it et al.}, Science {\bf 291}, 2570 (2001).

\bibitem{imambekov}
A. Imambekov and E. Demler; Phys. Rev. A {\bf 73}, 021602(R) (2006).

\bibitem{tkt}
K. Harada and N. Kawashima Phys. Rev. B {\bf 55}, R11949 (1998).

\bibitem{ceperley}
D. M. Ceperley, Rev. Mod. Phys. {\bf 67}, 279 (1995).

\bibitem{worldline}
J. E. Hirsch, R. L. Sugar, D. J. Scalapino and R. Blankenbecler,
Phys. Rev. B {\bf 26}, 5033 (1982).

\bibitem{prokofev}
N. V. Prokof\'ev, B. V. Svistunov, and I. S. Tupitsyn,
Zh. Eks. Teor. Fiz. {\bf 114}, 570 (1998) [JETP {\bf 87}, 311 (1998)].

\bibitem{directed}
O. F. Sylju{\aa}sen and A. W. Sandvik, Phys. Rev. E {\bf 66}, 046701 (2002). 

\bibitem{boninsegni}
M. Boninsegni, N. Prokof'ev, and B. Svistunov, 
Phys. Rev. Lett. {\bf 96}, 070601 (2006).

\bibitem{sse}
A. W. Sandvik and J. Kurkij{\"a}rvi, Phys. Rev. B {\bf 43}, 5950 (1991);
A. W. Sandvik, J. Phys. A {\bf 25}, 3667 (1992).

\bibitem{sseprb}
A. W. Sandvik, Phys. Rev. B {\bf 56}, 11678 (1997).

\bibitem{pollock}
E. L. Pollock and D. M. Ceperley, Phys. Rev. B {\bf 36}, 8343 (1987).

\end{thebibliography}
\end{document}